\documentclass{article}
\pdfoutput=1

\usepackage{arxiv}

\usepackage[utf8]{inputenc} 
\usepackage[T1]{fontenc}    
\usepackage{hyperref}       
\usepackage{url}            
\usepackage{booktabs}       
\usepackage{amsfonts}       
\usepackage{nicefrac}       
\usepackage{microtype}      
\usepackage{lipsum}		    
\usepackage[colorinlistoftodos,prependcaption]{todonotes}

\RequirePackage[]{natbib}

\usepackage[american]{babel}
\usepackage{csquotes}
\usepackage{amsmath}
\usepackage{todonotes}
\usepackage{tikz}
\usepackage{float}
\usepackage{lineno}
\usepackage{multirow}
\usepackage{adjustbox}
\usepackage{amsthm}
\usepackage{amssymb}
\usepackage{graphicx}
\usepackage[figuresright]{rotating}
\usepackage{comment}

\def\BF{\textsc{BF}}
\def\AF{\textsc{AF}}

\title{A general approximation to nested Bayes factors with informed priors}

\author{
    František Bartoš                    \\
	Department of Psychological Methods \\
	University of Amsterdam             \\
	Noord-Holland, The Netherlands      \\
	\And
	Eric-Jan Wagenmakers                \\
	Department of Psychological Methods \\
	University of Amsterdam             \\
	Noord-Holland, The Netherlands      \\
}

\renewcommand{\shorttitle}{Approximate informed Bayes factors}

\begin{document}
\maketitle

\begin{abstract}
A staple of Bayesian model comparison and hypothesis testing, Bayes factors are often used to quantify the relative predictive performance of two rival hypotheses. The computation of Bayes factors can be challenging, however, and this has contributed to the popularity of convenient approximations such as the BIC. Unfortunately, these approximations can fail in the case of informed prior distributions. Here we address this problem by outlining an approximation to informed Bayes factors for a focal parameter $\theta$. The approximation is computationally simple and requires only the maximum likelihood estimate $\hat\theta$ and its standard error. The approximation uses an estimated likelihood of $\theta$ and assumes that the posterior distribution for $\theta$ is unaffected by the choice of prior distribution for the nuisance parameters. The resulting Bayes factor for the null hypothesis $\mathcal{H}_0: \theta = \theta_0$ versus the alternative hypothesis $\mathcal{H}_1: \theta \sim g(\theta)$ is then easily obtained using the Savage--Dickey density ratio. Three real-data examples highlight the speed and closeness of the approximation compared to bridge sampling and Laplace's method. The proposed approximation facilitates Bayesian reanalyses of standard frequentist results, encourages application of Bayesian tests with informed priors, and alleviates the computational challenges that often frustrate both Bayesian sensitivity analyses and Bayes factor design analyses. The approximation is shown to suffer under small sample sizes and when the posterior distribution of the focal parameter is substantially influenced by the prior distributions on the nuisance parameters. The proposed methodology may also be used to approximate the posterior distribution for $\theta$ under $\mathcal{H}_1$. 
\end{abstract}


\section{Introduction}
Bayes factors represent the standard solution to problems involving Bayesian model comparison and hypothesis testing (e.g., \citealp{benjamin2018redefine, berger1987testing, johnson2023bayes, kass1995bayes}). Across the empirical sciences, the prototypical testing scenario features a null hypothesis $\mathcal{H}_0$ where a focal, test-relevant parameter $\theta$ is fixed to a particular value of interest: $\mathcal{H}_0: \theta = \theta_0$. The alternative hypothesis $\mathcal{H}_1$ relaxes the restriction imposed by $\mathcal{H}_0$. Here we consider the Bayesian framework in which the test-relevant parameter is assigned a prior distribution: $\mathcal{H}_1: \theta \sim g(\theta)$. Both $\mathcal{H}_0$ and $\mathcal{H}_1$ may additionally feature a common set of nuisance parameters $\psi$. The Bayes factor (e.g., \citealp{etz2017haldane,jeffreys1935some,jeffreys1939theory,kass1995bayes}) quantifies the evidence that the data $y$ provide for $\mathcal{H}_0$ versus $\mathcal{H}_1$ and is defined as the ratio of the two integrated likelihoods, that is, the ratio of the likelihoods integrated over the prior:
\begin{equation*}
    \BF_{10} = \frac{p(y \mid \mathcal{H}_1)}{p(y \mid \mathcal{H}_0)} = \frac{\int\int p(y \mid \theta, \psi, \mathcal{H}_1) \, p(\theta, \psi \mid \mathcal{H}_1) \, {\rm d} \psi \, {\rm d} \theta}{\int p(y \mid \theta = \theta_0, \psi, \mathcal{H}_0) \, p(\theta = \theta_0, \psi \mid \mathcal{H}_0) \, {\rm d} \psi}.
\end{equation*}
Hence the Bayes factor reflects the models' relative predictive performance, which also equals the extent to which the data warrant a change from prior to posterior model odds \citep{wrinch1921on}:
\begin{equation*}
    \BF_{10} = \frac{p(\mathcal{H}_1 \mid y)}{p(\mathcal{H}_0 \mid y)} \Bigg/ \frac{p(\mathcal{H}_1)}{p(\mathcal{H}_0)}.
\end{equation*}

Bayes factors allow researchers (1) to quantify evidence both for and against the null hypothesis \citep{jeffreys1935some,keysers2020using}; (2) to update evidence as the data accumulate, seamlessly and coherently \citep{edwards1963bayesian, berger1988likelihood}; and (3) to apply substantive background knowledge to increase the diagnosticity of the test (e.g., \citealp{johnson2009bayesian, johnson2010use, gronau2020informed}).

In most non-trivial applications, however, researchers who seek to obtain a Bayes factor are faced with considerable computational challenges. The integrals that define the integrated likelihood may be high dimensional, and state-of-the-art methods such as bridge sampling \citep{meng1996simulating, gronau2017tutorial} or reversible-jump Markov chain Monte Carlo \citep{green1995reversible} are generally time intensive. This concern is especially relevant for prior sensitivity analyses and Bayes factor design analyses that require repeated Bayes factor evaluations \citep{schonbrodt2018bayes, stefan2019tutorial}. In addition, models with many nuisance parameters are almost always applied using a default multivariate prior distribution that is difficult to adjust in light of substantive background knowledge concerning the focal, test-relevant parameter of interest.

These challenges are often side-stepped by convenient approximations to the integrated likelihood such as the Bayesian information criterion (BIC; \citealp{schwarz1978estimating}) and Laplace's method \citep{tierney1986accurate}, or an approximation to default Bayes factors requiring only sample size and a test-statistic or $p$-value (\citealp{jeffreys1936further, wagenmakers2022approximate}, also see \citealp{johnson2005bayes, shao2019bayesian, villa2022objective, rostgaard2023simple} for other approaches). However, these approximations have notable limitations.

First, the BIC is defined as $-2\log{p(y \mid \hat{\xi})} + k\log{n}$, that is, the maximum likelihood plus a complexity correction term that contains the number of free parameters $k$ and the sample size $n$. Unfortunately, both $k$ and $n$ can be surprisingly difficult to determine \citep{kass1995bayes, pauler1998schwarz}. Moreover, the BIC approximates a ``default'' Bayes factor that is based on a unit-information prior \citep{kass1995reference}; consequently, the BIC does not easily lend itself to an analysis that seeks to take advantage of substantial background knowledge. 

Second, Laplace's method assumes that the posterior distribution is highly peaked around the maximum likelihood estimate and approximates the integrated likelihood under each hypothesis $\mathcal{H}_{.}$ as $(2\pi)^{k/2}\mid\hat{\Sigma}\mid^{1/2}p(y \mid \hat{\theta},\hat{\psi},\mathcal{H}_{.}) \, p(\hat{\theta},\hat{\psi}\,\mid\,\mathcal{H}_{.})$, where $\hat{\Sigma}$ is the covariance matrix of the $k$ maximum likelihood estimates $\hat{\theta},\hat{\psi}$ \citep{kass1995bayes}. Laplace's method effectively assumes that the posterior distribution is multivariate normal, fully determined by the likelihood function, and not influenced by the shape of the prior distributions. Consequently, Laplace's method can perform poorly with informed prior distributions. 

Finally, Jeffreys's approximate Bayes factor (first mentioned in \citealp[p. 417]{jeffreys1936further}) assumes that the prior distribution $\theta \sim g()$ varies slowly in the neighborhood of the maximum likelihood estimate and provides a convenient test against a null hypothesis $\mathcal{H}_{0}: \theta = 0$. Jeffreys's approximate Bayes factor simplifies to $\BF_{01} = A \sqrt{n} \, \text{exp}(-\chi^2/2)$, where $\chi^2$ corresponds to a Wald statistic and $A$ is a constant usually close to 1 (see \citealp{wagenmakers2022approximate} for an overview). Although the prior sensitivity of Jeffreys's approximate Bayes factor can to some degree be accommodated by adjusting the $A$ argument, the general expression with $A=1$ corresponds to an ``objective'' unit-information Bayes factor that does not test informed hypotheses. Other approaches (e.g., \citealp{johnson2005bayes, shao2019bayesian, villa2022objective, rostgaard2023simple}) also focus on Bayes factors under ``objective'' or improper prior distributions which we do not consider in this manuscript. 

In sum, all three approximate methods outlined above are appropriate only for scenarios with relatively uninformed prior distributions \citep{kass1995bayes}. Consequently, accurate approximations are lacking for exactly the type of testing scenario in which Bayesian inference ought to excel: the case where substantial prior knowledge is available \cite{gronau2020informed}.

To overcome this limitation we outline a simple method that can approximate informed Bayes factors for focal parameters in nested models differing only in the presence or absence of a single test-relevant parameter $\theta$. The method takes advantage of the approximate likelihood function of \citet{tsou1995robust} and the Savage--Dickey density ratio \citep{dickey1971weighted, dickey1970weighted, wetzels2009quantify}, the same principle recently used by \cite{johnson2023bayes, mulder2020generalization, rostgaard2023simple}.\footnote{The results are equivalent to Bayes factor $z$-tests for maximum likelihood parameter estimates $\hat{\theta}$ with known standard errors $\text{se}(\hat{\theta})$, as in \citet{spiegelhalter2004bayesian}, \citet{dienes2014using}, and \citet{dienes2018four}.} The approximation requires a maximum likelihood estimate $\hat{\theta}$, its standard error $\text{se}(\hat{\theta})$, and holds under weak regularity conditions. Appendix 1 shows how the resulting Bayes factor can be readily computed in \texttt{R} or JASP \citep{R, JASP16, ly2021bayesian}.

We refer to the proposed methodology as the \emph{Savage--Dickey normal approximation} and illustrate its precision and utility with three real-data examples. The first example features a two-sample $t$-test and compares the results to those obtained using standard numerical methods. The second example features a sequential parametric survival analysis and compares the results to those obtained using bridge sampling and Laplace's method. The third example features meta-regression and presents a comparison based on a prior sensitivity analysis. The closing section points to limitations and outlines further advantages of the approximation.

\section{Methods}
As the name implies, the Savage--Dickey normal approximation is based on the Savage--Dickey density ratio which obviates the need to compute the ratio of two marginal likelihoods; instead, it expresses the Bayes factor $\BF_{10}$ for $\mathcal{H}_1: \theta \sim g(\theta)$ against a point null hypothesis $\mathcal{H}_0: \theta = \theta_0$ as a ratio of the prior ordinate over the posterior ordinate under $\mathcal{H}_1$ evaluated at the test value $\theta_0$:
\begin{equation*}
\label{eq:savage-dickey}
\BF_{10} = \frac{p(\theta = \theta_0 \mid \mathcal{H}_1)}{p(\theta = \theta_0 \mid y, \mathcal{H}_1)}.
\end{equation*}
The Savage--Dickey density ratio assumes that $p(\psi  \mid \mathcal{H}_0) = p(\psi \mid \theta = \theta_0,\mathcal{H}_1)$, that is, the prior distributions on the nuisance parameters are specified such that $\mathcal{H}_1$ reduces to $\mathcal{H}_0$ when the focal parameter equals $\theta_0$ in both models (\citealp[p. 249]{jeffreys1961theory}; \citealp{verdinelli1995computing}; for a generalization that relaxes this assumption see \citealp{verdinelli1995computing, heck2019caveat, mulder2020generalization}).

The prior ordinate for $\theta$ under $\mathcal{H}_1$ evaluated at $\theta_0$ is available directly from the prior probability density function, but the marginal posterior density function $p(\theta \mid y, \mathcal{H}_1)$ usually does not have a closed-form solution. Therefore, we obtain the posterior ordinate for $\theta$ at $\theta_0$ using an approximate marginal posterior probability density function $p_a(\theta \mid y, \mathcal{H}_1)$.

We construct this approximate marginal posterior probability density function via the approximate likelihood function 
\begin{equation*}
L_a(\theta) = \text{exp}\left(-\tfrac{1}{2}\frac{(\hat{\theta} - \theta)^2}{\text{se}(\hat{\theta})^2}\right),
\end{equation*}
which is proportional to a normal density. \citet{tsou1995robust} showed that the estimated likelihood $L_a(\theta) = L(\theta, \hat{\psi})$ is asymptotically globally pointwise equivalent to the complete likelihood $L(\theta, \psi)$ under weak regularity conditions.\footnote{Global pointwise equivalence holds when the parameter space of $\psi$ is discrete or --in case the parameter space of $\psi$ is continuous-- the likelihood $l(\theta, \psi)$ has a continuous derivative with respect to $\psi$ at $\hat{\psi}$ and $\hat{\psi}$ is a consistent estimator of $\psi$ \citep{tsou1995robust}.} In other words, $L_a(\theta)$ is asymptotically equivalent to $L(\theta, \psi)$ when comparing support provided by the data between any two values of $\theta$ via likelihood ratios (\citealp[p. 158]{royall1997statistical}). Next we apply Bayes' rule to obtain the approximate marginal posterior distribution for $\theta$,
\begin{equation*}
p_a(\theta \mid  y, \mathcal{H}_1) = \frac{L_a(\theta) \, p(\theta \mid \mathcal{H}_1)}{\int L_a(\theta) \, p(\theta \mid \mathcal{H}_1) \, {\rm d} \theta},
\end{equation*}
as a standardized product of the approximate likelihood of $\theta$ and the prior distribution of $\theta$ (see \citealp{pratt1965bayesian} for the same approximation with insufficient statistics). Since we need to approximate only the \emph{marginal} posterior distribution of $\theta$, the denominator features only a one-dimensional integral that can easily be evaluated numerically. This can be considered a simple extension of the Laplace approximation to the posterior distribution \citep{leonard1982comment} that takes into account the prior information without assuming the posterior is normally distributed.

Finally, we substitute the approximate posterior distribution into the Savage--Dickey expression for the Bayes factor and obtain the Savage--Dickey normal approximation as follows:
\begin{equation*}
\label{eq:savage-dickey-approx}
\BF_{10} \approx \frac{\int L_a(\theta) \, p(\theta \mid \mathcal{H}_1) \, {\rm d} \theta}{L_a(\theta_0)}.
\end{equation*}

When the focal parameter $\theta$ is assigned a normal prior distribution with mean $\mu_0$ and standard deviation $\sigma_0$, $\mathcal{H}_1: \theta \sim N(\mu_0,\sigma_0^2)$, the Savage--Dickey normal approximation conveniently yields a closed-form expression for the Bayes factor:
\begin{equation*}
\label{eq:normal-normal}
\BF_{01} \approx \sqrt{\frac{ \sigma_0^2  + \text{se}(\hat{\theta})^2 }{\text{se}(\hat{\theta})^2}} \,\, \exp\left( -\tfrac{1}{2} \left[ \frac{(\hat{\theta} - \theta_0)^2}{\text{se}(\hat{\theta})^2} - \frac{(\hat{\theta} - \mu_0)^2}{\sigma_0^2 + \text{se}(\hat{\theta})^2} \right] \right)
\end{equation*}

Appendix~2 shows that this expression, which corresponds to a Bayesian $z$-test (cf. \citealp{berger1987testing}, Eq.~6, and \citealp{clyde2021introduction}), approaches Jeffreys's default approximate Bayes factor (e.g., \citealp[p. 247]{jeffreys1961theory}) with increasing sample size and under a unit information prior; as an approximate Bayes factor for logistic regression the expression has been advocated by \citet{wakefield2007bayesian, wakefield2009bayes}.

\section{Examples}

\subsection{Two-sample $t$-test}

\begin{figure}[h]
    \centering
    \includegraphics[width=.50\textwidth]{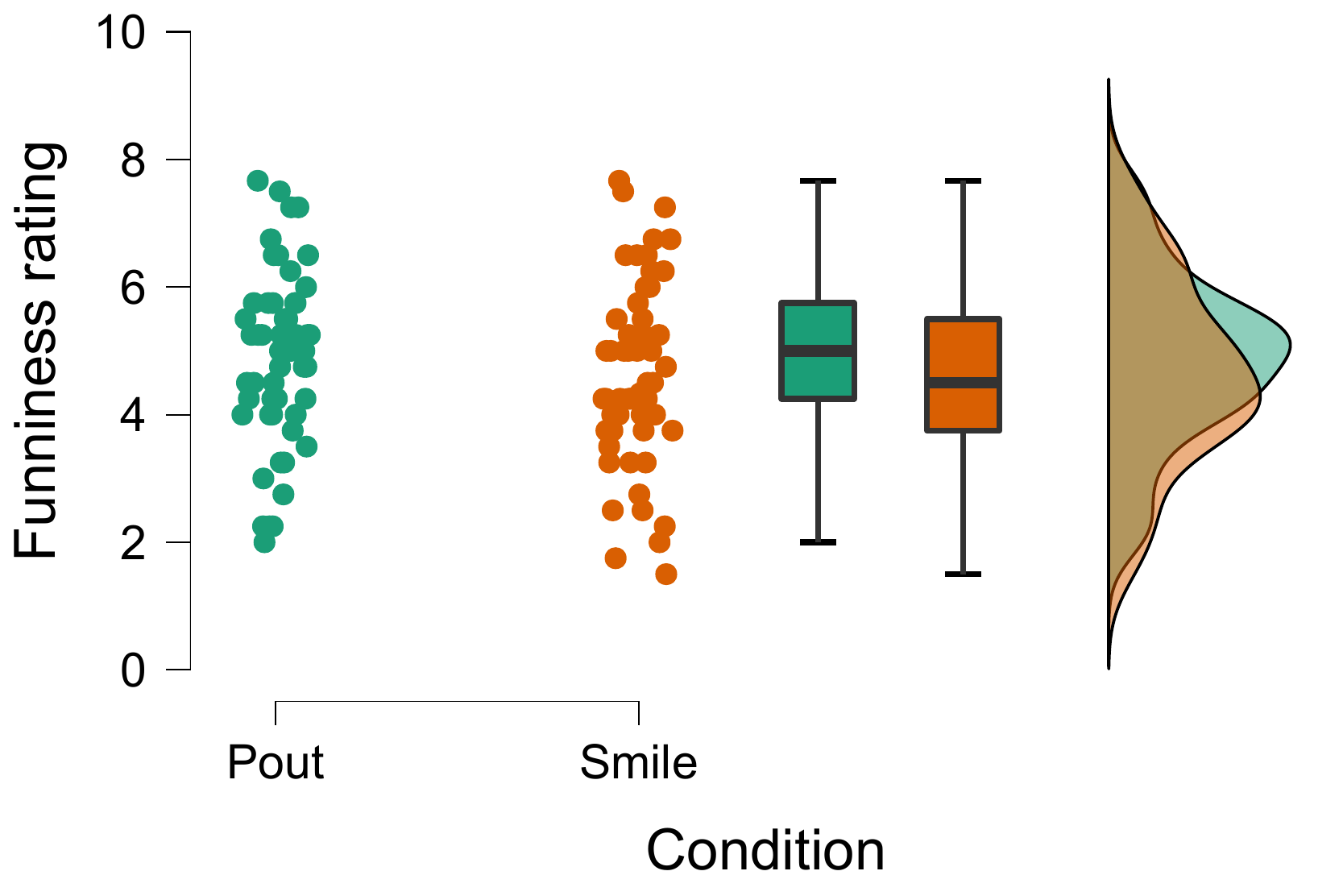}
    \caption{Distribution of mean cartoon funniness ratings in the pout and smile condition. Data from the Oosterwijk's laboratory in the replications study of a facial feedback hypotheses by \citeauthor{wagenmakers2016registered}. Figure from JASP.}
    \label{fig:t-test-rating}
\end{figure}

First we compare the Savage--Dickey normal approximation to a numerical solution for the informed Bayesian two-sample $t$-test  \citep{gronau2020informed}. We consider \citeauthor{gronau2020informed}'s re-analysis of a replication study of the facial feedback hypothesis \citep{wagenmakers2016registered}, which holds that facial expressions can impact emotional experience. In the replication study, participants were instructed to rate the funniness of a cartoon while holding a pen either with their teeth, i.e., the smile condition, or with their lips, i.e., the pout condition. The facial feedback hypothesis predicts that participants in the the smile condition will rate the cartoons to be funnier than participants in the pout condition (Figure~\ref{fig:t-test-rating}).

In the replication study, \citet{wagenmakers2016registered} found a mean funniness rating of $4.63$ ($\text{SD} = 1.48$) across $53$ participants in the smile condition and $4.87$ ($\text{SD} = 1.32$) across $57$ participants in the pout condition. To specify the prior distribution on the effect size parameter $\delta$, \citet{gronau2020informed} performed prior elicitation with an expert from the field of social psychology and obtained an informed Student-$t$ prior distribution, called the ``Oosterwijk'' prior distribution after the expert. The prior distribution specifies mostly small effect sizes and is restricted to positive values: $\delta \sim \text{Student-t}_{+}(\mu = 0.350, \sigma = 0.102,\nu = 3)$.

The numerical solution presented by \citet{gronau2020informed} shows strong evidence in favor of the null hypothesis, $\BF_{0+} = 11.6$. The effect in the sample is in the direction opposite to that predicted by the facial feedback hypothesis.\footnote{The $0.1$ difference between the $\BF_{01}$ reported here and in \citet{gronau2020informed} is due to rounding.} The Savage--Dickey normal approximation with the maximum likelihood estimate $\hat{\delta} = -0.17$ and standard error $\text{se}(\hat{\delta}) = 0.19$ leads to almost identical evidence in favor of the null hypothesis: $\BF_{0+} = 11.5$.

We found that the Savage--Dickey normal approximation Bayes factor closely corresponds to the numerical solution by \citet{gronau2020informed} for all reasonable effect sizes $\mid \delta \mid < 1$. Larger sample sizes and effect sizes can result in notable underestimation of the evidence in the favor of the alternative hypothesis $\mathcal{H}_1$. However, the evidence in favor of $\mathcal{H}_1$ is already so large, e.g., $\BF_{10} > 10^{100}$, that a ten times lower Bayes factor obtained by the Savage--Dickey normal approximation does not change the qualitative assessment of evidence.\footnote{The performance of the Savage--Dickey normal approximation could be further improved by reparametrizing the approximation in the terms of a $z$-statistic and transforming a $p$-value obtained from the $t$-test into a $z$-statistic. This modification corrects for a higher ordinate at $p_a(0 \mid y, \mathcal{H}_1)$ when the maximum likelihood estimate $\hat{\theta}$ is very far away from zero which is introduced by the slightly skewed likelihood of the $t$-test.}

\subsection{Parametric survival analysis}

Next we compare the Savage--Dickey normal approximation to Laplace's method and bridge sampling for an informed Bayesian parametric survival analysis \citep{bartos2021informed}. We repeat \citeauthor{bartos2021informed}'s full sample and sequential re-analysis of a colon cancer treatment trial that examined the potential increase in disease-free survival due to adding Cetuximab to the standard sixth version of a FOLFOX regimen \citep{alberts2012coloncancer}.

The data set obtained from Project Data Sphere \citep{re3data2019project} contains $22.9$\% recurrences across 1247 participants in the standard treatment group and $22.9$\% recurrences across 1251 participants in the enhanced treatment group. We perform two analyses: (1) we specify an uninformed standard normal distribution on the log acceleration factors, $\text{log}(\AF) \sim \text{Normal}(0, 1)$, and (2) we specify and informed directional hypotheses test, $\text{log}(\AF) \sim \text{Normal}_+(0.30, 0.15^2)$, as performed by \citet{bartos2021informed}. Laplace's method should perform relatively well in the first scenario, with weak prior information, and relatively poorly in the second scenario, with strong prior information. 

We focus solely on log-normal survival model that received the highest posterior model probability in the re-analysis. For the nuisance parameters, we use informed prior distributions as specified in Table~1 in \citet{bartos2021informed}.

\begin{figure}[h]
    \centering
    \includegraphics[width=.9\textwidth]{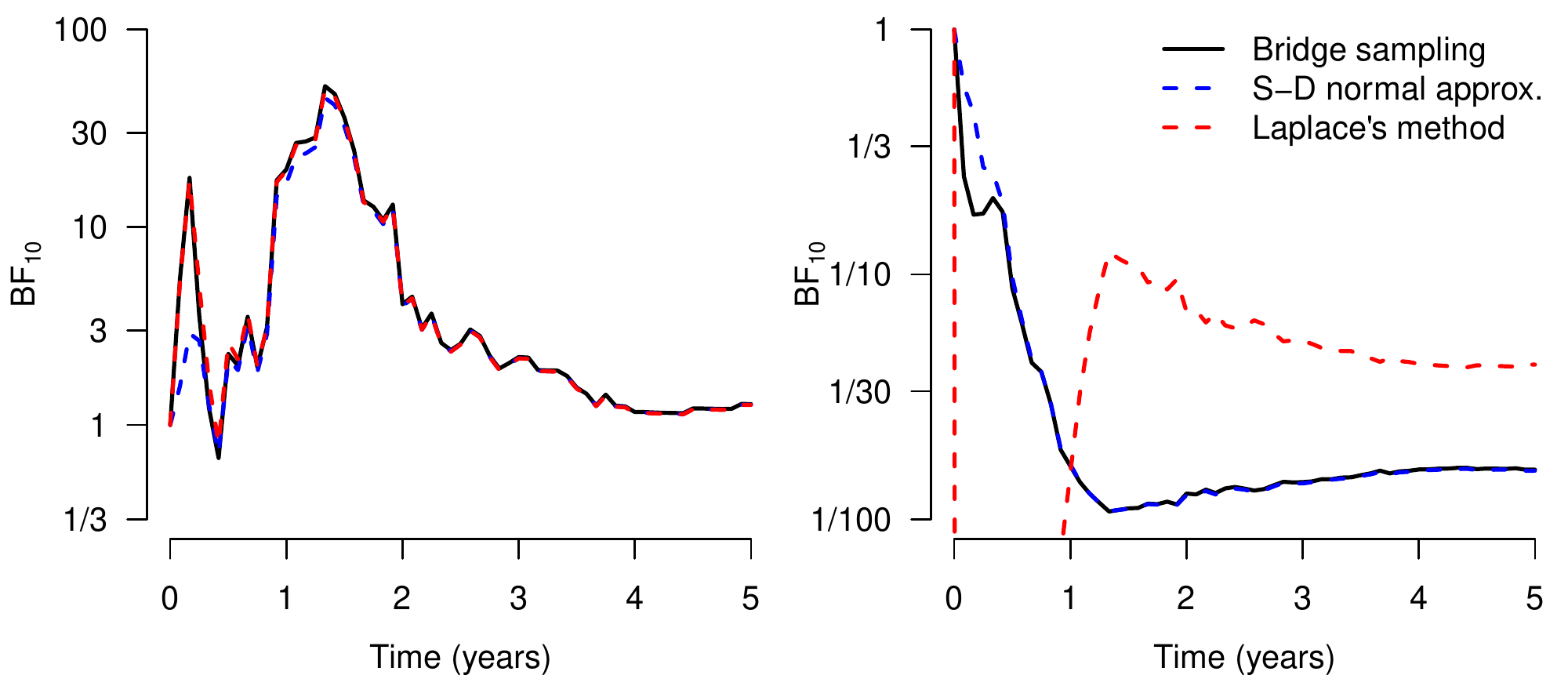}
    \caption{Comparison of different methods for computing Bayes factors for a sequential analysis of the survival analysis example. The solid black line corresponds to bridge sampling, representing the gold standard, the dashed blue line corresponds to the Savage--Dickey normal approximation, and the dashed red line corresponds to Laplace's method. Left panel: Bayes factor for the treatment effect, i.e., the log acceleration factor, under a weakly informed $\text{Normal}(0, 1)$ prior distribution. Right panel: Bayes factor for the treatment effect, i.e., the log acceleration factor, under an informed $\text{Normal}_{+}(0.30, 0.15^2)$ prior distribution. Laplace's method would be undefined for the right panel since the maximum likelihood estimate of the log acceleration factor is negative; therefore we modified the procedure by ignoring the truncation of the prior distribution.}
    \label{fig:survival-sequential}
\end{figure}

In the scenario with the standard normal prior distribution, i.e., weak prior information, the precise Bayes factor computed by bridge sampling on the complete data set shows an absence of evidence, $\BF_{10} = 1.3$, as does the Savage--Dickey normal approximation, $\BF_{10} = 1.3$, and Laplace's method, $\BF_{10} = 1.3$. The left panel of Figure~\ref{fig:survival-sequential} compares the results of both approximations to bridge sampling when the data are analyzed as they accumulate over time. We see that especially early in the trial, when the number of observed events is low (i.e., 8 vs. 1, and 13 vs. 2 events in the experimental and control conditions, respectively), Laplace's method approximates the precise Bayes factor better than the Savage-Dickey normal approximation, possibly due to the impact of the informed prior distributions on the nuisance parameters which is not accounted for by the Savage--Dickey normal approximation. Nevertheless, the Savage--Dickey normal approximation quickly converges to the precise Bayes factor as well.

Using the informed directional hypothesis test illustrates one of the limitations of Laplace's method; at the conclusion of the trial the maximum likelihood estimate of the log acceleration factor is negative, $\text{log}(\widehat{\AF}) = -0.19$, $\text{se}(\text{log}(\widehat{\AF})) = 0.08$, as it is throughout most of its sequential trajectory. In order to obtain an approximation using Laplace's method at all, we removed the lower truncation from its computation (the maximum likelihood estimate is not inside the region of interest). Using the complete data set, the precise computation using bridge sampling shows strong evidence in favour of the absence of the treatment effect, $\BF_{0+} = 61.8$, a value that is closely approximated by the Savage--Dickey normal approximation, $\BF_{0+} = 63.2$, but poorly approximated by the modified Laplace's method, $\BF_{0+} = 23.4$. The right panel of Fig.~\ref{fig:survival-sequential} tells a similar story when accumulating the evidence over time; the Savage--Dickey normal approximation quickly converges to the precise computation with increasing number of observed events, whereas the modified Laplace's method is unable to provide the expected answer.

The sequential analysis also illustrates the practical utility of the Savage--Dickey normal approximation for Bayes factor design analyses, in which hundreds of Bayes factor trajectories need to be computed under different simulated data sets. For this single Bayes factor trajectory consisting of 60 Bayes factors, the computation required $3.4$ and $3.5$ CPU hours for the bridge sampling implementation, $0.1$ and $0.4$ CPU \emph{seconds} for the Savage--Dickey normal approximation, and $8.2$ and $7.7$ CPU \emph{seconds} for Laplace's method, for the uninformed and informed specifications, respectively.\footnote{This and the following computations were done on a modern computer with AMD Ryzen 3900X CPU. The posterior samples for bridge sampling were estimated with a JAGS model with two chains, each running for 1000 adaptation, 1000 burn-in, and 4000 sampling iterations. See the Supplementary Materials at \url{https://osf.io/8vncp/} for more details.}

\subsection{Meta-Regression}

Finally we compare the Savage--Dickey normal approximation to Laplace's method and bridge sampling for a meta-regression analysis. In the original article, \citet{gronau2017powerpose} combined evidence across six replication studies concerning the (weak form of) power posing hypothesis, stating that an expansive body posture can increase the subjective feeling of power. We extend the analysis by testing and adjusting for a moderator: participants' self-reported familiarity with the power posing hypothesis.

\begin{table}
\centering
\caption{Effect sizes and standard errors of the effect of power posing on perceived power across six replication studies split by participants' familiarity with the power posing hypothesis}{
\begin{tabular}{lcccc}
                            & \multicolumn{2}{c}{Familiar} & \multicolumn{2}{c}{Non-familiar} \\
  Study                     & Effect size & Standard error & Effect size & Standard error     \\ 
  \hline
  Bailey et al. (2017)      & 0.05        & 0.77           & 0.24        & 0.40               \\ 
  Ronay et al. (2017)       & 0.16        & 0.31           & 0.21        & 0.48               \\ 
  Klaschinski et al. (2017) & 0.33        & 0.22           & 0.16        & 0.31               \\ 
  Bombari et al. (2017)     & 0.38        & 0.69           & 0.15        & 0.48               \\ 
  Latu et al. (2017)        & 0.16        & 0.11           & 0.15        & 0.42               \\ 
  Keller et al. (2017)      & 0.03        & 0.28           & 0.17        & 0.18               \\ 
\end{tabular}}
\label{tab:power-pose-data}
\end{table}

Table~\ref{tab:power-pose-data} summarizes the effect size estimates $y$ and their standard errors $\text{se}(y)$ split by participants' familiarity with the power posing hypothesis. We specify a fixed-effect meta-regression model, $y \sim \text{Normal}(\alpha + \beta x, \,  \text{se}(y)^2)$, with the intercept, $\alpha$, corresponding to the overall (unweighted) mean effect size and a moderator, $\beta$, accounting for the difference based on participants' self-reported familiarity ($x$; yes = $0.5$, no = $-0.5$) with the power posing hypothesis. We use default independent Cauchy prior distributions for both parameters, $\alpha, \beta \sim \text{Cauchy}(0, 1/\sqrt{2})$, \citep{BayesFactor} and perform prior sensitivity analysis for scale parameters of the prior distributions. 

Using the default prior distribution, the precise Bayes factor computed by bridge sampling shows strong evidence for the presence of the overall effect of power posing on perceived power, $\BF_{10}^{\alpha} = 88.0$, and moderate evidence against moderation by participants' familiarity with the power posing hypothesis, $\BF_{10}^{\beta} = 0.22$. Essentially equivalent results are obtained by the Savage--Dickey normal approximation, $\BF_{10}^{\alpha} = 87.9$ and $\BF_{10}^{\beta} = 0.23$, and by Laplace's method, $\BF_{10}^{\alpha} = 89.4$, $\BF_{10}^{\beta} = 0.23$.

\begin{figure}[h]
    \centering
    \includegraphics[width=.9\textwidth]{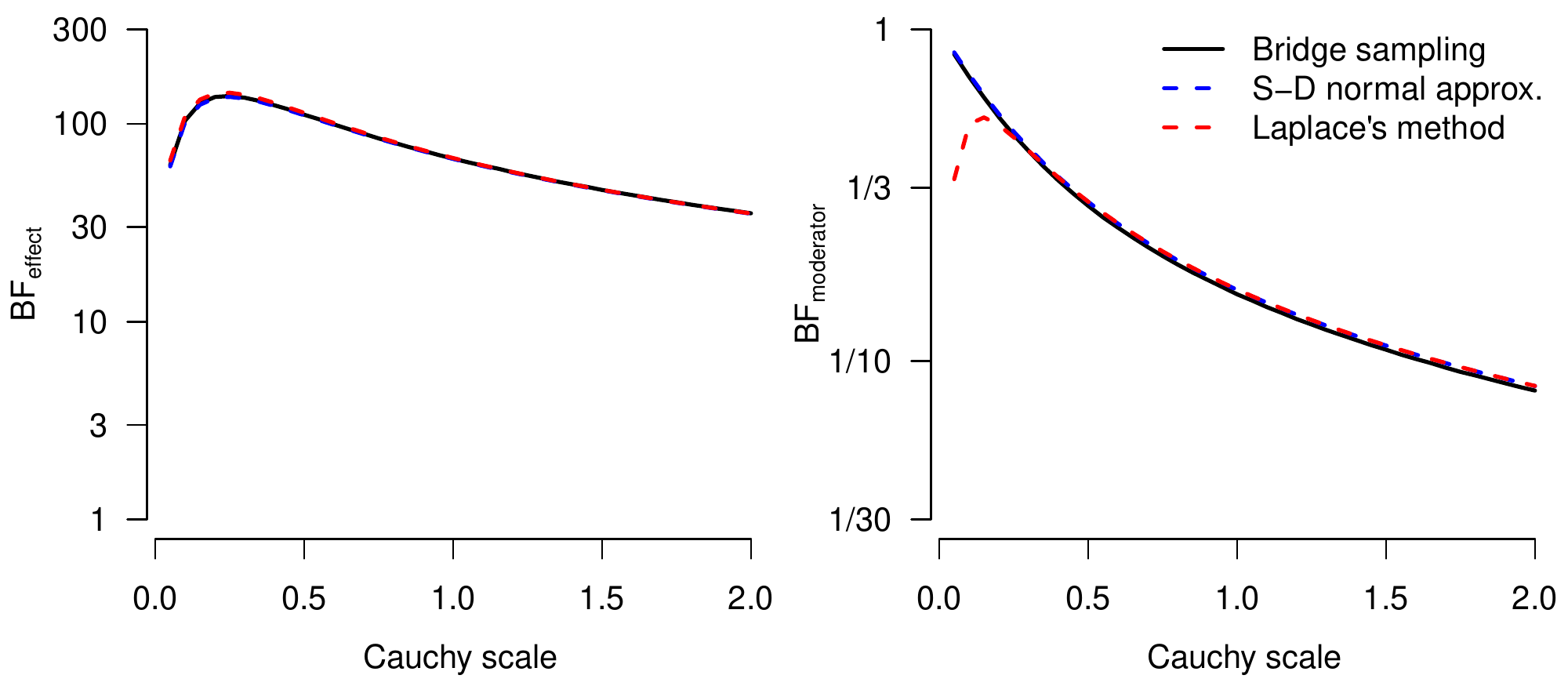}
    \caption{Comparison of different methods for computing Bayes factors for prior sensitivity analysis of the meta-regression example. The solid black line corresponds to bridge sampling, representing the gold standard, the dashed blue line corresponds to the Savage--Dickey normal approximation, and the dashed red line corresponds to Laplace's method. Left panel: Bayes factor for the overall meta-analytic mean effect with varying scale of the prior Cauchy distribution. Right panel: Bayes factor for the moderation by participants' familiarity with the power posing hypothesis with varying scale of the prior Cauchy distribution.}
    \label{fig:meta-regression-sensitivity}
\end{figure}

Figure~\ref{fig:meta-regression-sensitivity} compares results of both approximations to bridge sampling for the prior sensitivity analysis for the effect of power posing on perceived power (left panel) and the moderation by participants' familiarity (right panel). For the overall effect of power posing the results of all three methods are virtually indistinguishable. For the moderation by participants' familiarity, however, Laplace's method performs poorly for small scales of the Cauchy prior distribution -- that is, for informed prior distributions.

The prior sensitivity analysis again demonstrates the practical utility of the Savage--Dickey normal approximation. Computation of 40 Bayes factors for each parameter took $52.1$ CPU seconds for bridge sampling, which is substantially more demanding than the computation of the Savage--Dickey approximation, $0.24$ CPU seconds, or Laplace's method, $0.05$ CPU seconds. The speed difference was compounded by the fact that whereas the Savage--Dickey normal approximation requires only the estimation of a single model yielding both maximum likelihood estimates and multiplying each of them with their prior distributions, the bridge sampling implementation requires the estimation of three Stan models.

\section{Concluding Comments}

Popular approximations to the Bayes factor hypothesis tests such as BIC and Laplace's method fail in the case of informed prior distributions---arguably exactly the kind of scenario in which the demand for a Bayesian method is most acute. To overcome this limitation we outlined the Savage--Dickey normal approximation. This approximation is simple, accurate, and fast. Compared to the commonly used Laplace's method or BIC, the approximation allows researchers to test informed hypotheses instead of forcing them to perform different tests that they intended, or even relying on more ad hoc methods. Compared to the gold-standard bridge sampling implementation, the approximation is several orders of magnitude faster, a reduction of computation time that is especially useful in Bayesian design analyses and prior sensitivity analyses. Furthermore, because the maximum likelihood estimate and its standard error are commonly reported, the Savage--Dickey normal approximation presents a particularly straightforward method to re-evaluate claims from the literature. The approximation also provides a sanity check for researchers wishing to implement precise Bayes factor alternatives to existing frequentist tests.

The Savage--Dickey normal approximation is not without limitations. Specifically, the approximation does not apply to non-nested model comparison \citep[e.g., different parametric families as in][]{bartos2021informed}. Also, the approximation assumes that the prior distributions on the nuisance parameters do not strongly impact the posterior distribution of the focal parameter. Even though this is often the case \citep[pp. 249-251]{jeffreys1961theory}, there exist several applications of Bayesian inference with restrictive prior distributions on nuisance parameters to help regularize the estimates for focal parameter \citep[e.g., ][]{maier2020robust, bartos2021no}. In some cases, this limitation might be addressed by incorporating the restriction into the likelihood directly.

Finally, throughout the manuscript we used a normal distribution to approximate the likelihood. This might not always lead to acceptable results; in some cases, a log or logistic transformation of the focal parameter is advisable \citep{leonard1982comment, mulder2020generalization}. In other cases, a completely different likelihood function might be used directly, e.g., binomial or Student-$t$. We also observed that the Savage--Dickey normal approximation can lose precision with increasing distance of the maximum likelihood estimate from the test value $\theta_0$. This is a consequence of a decreasing appropriateness of the approximate likelihood function for the extreme tails. However, in such cases the data are so diagnostic that the results will pass the ``interocular traumatic test'', attributed to Berkson ( \citealp{edwards1963bayesian}). The approximation still leads to a qualitatively same conclusion, although a more precise methods should be used if high precision results are demanded.

Some of the above limitations may be re-phrased as advantages; for instance, the approximation does not require the analyst to specify prior distributions on nuisance parameters, which can be challenging. Furthermore, since the approximation concerns a single focal parameter, it does not fall prey to the Borel–Kolmogorov paradox \citep{consonni2008compatibility}. We believe that the Savage--Dickey normal approximation provides a straightforward and attractive alternative to other ways of approximating Bayes factors for nested models, especially in cases with informed prior distributions on the parameter of interest.

\section*{Acknowledgement}

This work was supported by The Netherlands Organisation for Scientific Research (NWO) through a Vici grant and an Advanced ERC grant (743086 UNIFY) to Eric-Jan Wagenmakers. This publication is based on research using information obtained from \url{www.projectdatasphere.org}, which is maintained by Project Data Sphere. Neither Project Data Sphere nor the owner(s) of any information from the web site have contributed to, approved or are in any way responsible for the contents of this publication.
We are thankful to Samuel Pawel for helpful comments and suggestions on an earlier version of the manuscript.

\section*{Supplementary Material}

The analysis scripts are available at \url{https://osf.io/8vncp/}. The data required for the two-sample $t$-test and the meta-regression example are also available at \url{https://osf.io/8vncp/}. The data required for the parametric survival analysis example are available from \url{www.projectdatasphere.org} upon a simple registration.

\section*{Competing Interests}
The authors declare no competing interests.

\appendix

\section*{Appendix 1}
\label{app:computation}
\subsection*{Computing the Savage--Dickey approximation with \texttt{R}}

The proposed approximation can be easily obtained through the \texttt{bayesplay} \texttt{R} package \citep{bayesplay}. Here we reproduce the results for Example 1: the two-sample $t$-test. We load the package and specify the prior distribution, $\delta \sim \text{Student-t}_{+}(\mu = 0.350, \sigma = 0.102,\nu = 3)$, and likelihood  $p(y \mid \delta) = \text{Normal}(-0.17, 0.19^2)$ as follows:
\begin{verbatim}
library(bayesplay)
prior      <- prior("student_t", mean = 0.350, sd = 0.102, df = 3,
                    range = c(0, Inf))
likelihood <- likelihood("normal", mean = -0.17, sd = 0.19)
\end{verbatim}
Next we obtain the posterior distribution and compute the Savage--Dickey approximation:
\begin{verbatim}
posterior  <- prior * likelihood
prior$prior_function(0) / posterior$posterior_function(0)
> 0.08585957
\end{verbatim}

\subsection*{Computing the Savage--Dickey approximation with JASP}

The proposed approximation can also be computed using JASP \citep{JASP16, ly2021bayesian}. JASP provides a graphical user interface for the \texttt{bayesplay} \texttt{R} package \citep{bayesplay} in the ``General Bayesian Tests'' analysis within the ``Summary Statistics'' module (that must be enabled by clicking the large blue `$+$' sign in the upper right corner of the JASP screen).

\begin{figure}[h]
    \includegraphics[width=\textwidth]{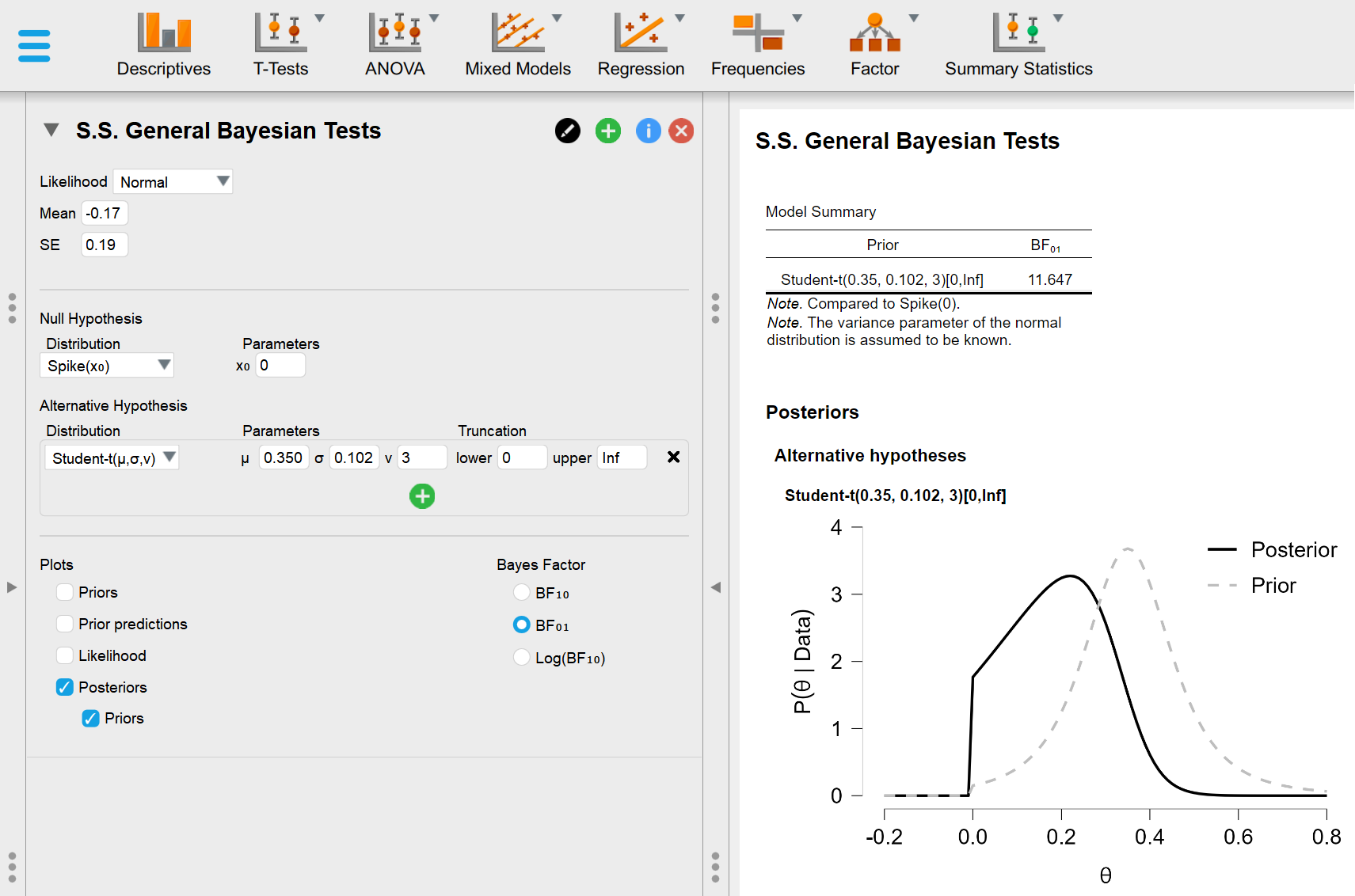}
    \caption{JASP implementation of the Savage--Dickey normal approximation for Example 1: the two-sample $t$-test. The left panel of the graphical user interface contains the input, that is, the likelihood and the prior distribution of the null and alternative hypothesis. The right panel presents the output, that is, the Bayes factor in favour of the null hypothesis over the informed directional alternative hypothesis. Screenshot from JASP.}
    \label{fig:JASP}
\end{figure}

Figure~\ref{fig:JASP} shows a JASP screenshot; the left panel of the graphical user interface contains the specification of the two-sample $t$-test from Example 1, whereas the right panel presents the results.

\section*{Appendix 2}
\label{app:Jeffreys}

\subsection*{Normal-normal approximation}

Assume the approximate likelihood $L_a(\theta)$ proportional to a normal density with mean equal to the maximum likelihood estimate $\hat{\theta}$ and standard deviation equal to the standard error of the maximum likelihood estimate $\text{se}(\hat{\theta})$. When the focal parameter $\theta$ is assigned a normal prior distribution with mean $\mu_0$ and standard deviation $\sigma_0$, $\theta \sim \text{Normal}(\mu_0, \sigma_0^2)$ under $\mathcal{H}_1$ and $\theta = \theta_0$ under $\mathcal{H}_0$, the Savage--Dickey density ratio representation of the Bayes factor,

\begin{equation*}
\BF_{01} = \frac{p(\theta = \theta_0 \mid y, \mathcal{H}_1)}{p(\theta = \theta_0 \mid \mathcal{H}_1)},
\end{equation*}

is approximated by

\begin{equation*}
\BF_{01} \approx \frac{N\left(\theta_0 \mid \frac{\sigma_0^2 \hat{\theta} + \text{se}(\hat{\theta})^2 \mu_0}{\sigma_0^2 + \text{se}(\hat{\theta})^2}, \frac{\sigma_0^2 \text{se}(\hat{\theta})^2}{\sigma_0^2  + \text{se}(\hat{\theta})^2}\right)}{N(\theta_0 \mid \mu_0, \sigma_0^2)},
\end{equation*}

which simplifies to

\begin{equation*}
\BF_{01} \approx \sqrt{\frac{\sigma_0^2}{\text{se}(\hat{\theta})^2} + 1} \,\, \exp\left( -\tfrac{1}{2} \left[ \frac{(\hat{\theta} - \theta_0)^2}{\text{se}(\hat{\theta})^2} - \frac{(\hat{\theta} - \mu_0)^2}{\sigma_0^2 + \text{se}(\hat{\theta})^2} \right] \right)
\end{equation*}

\subsection*{Asymptotic equivalence to Jeffrey's general approximate Bayes factor}
When the focal parameter $\theta$ is assigned a unit information prior distribution, $\theta \sim \text{Normal}(\hat{\theta}, n \, \text{se}(\hat{\theta})^2)$, and we test against the null hypothesis $\theta_0 = 0$, substitution into the Normal-Normal approximation yields 

\begin{equation*}
\BF_{01} \approx \sqrt{n+1} \exp{\left(-\tfrac{1}{2} \frac{\hat{\theta}^2}{\text{se}(\hat{\theta})^2}\right) }.
\end{equation*}

With increasing sample size, this expression approaches Jeffreys's general approximate Bayes factor,
\begin{equation*}
\BF_{01} = A \sqrt{n} \exp{\left(-\tfrac{1}{2}  \frac{\hat{\theta}^2}{\text{se}(\hat{\theta})^2}\right) },
\end{equation*}
where $A$ is ``usually not far from 1'' \citep[p. 89]{jeffreys1977probability}.

\bibliographystyle{biometrika}
\bibliography{manuscript.bib}

\end{document}